# Magnet design optimization of a 100 MeV separated sector medical cyclotron and its injection line

Ali Biganeh[1], Ali Ramazani Moghaddam[1]

[1]Department of Energy Engineering & Physics, Amir Kabir University of Technology, Tehran, Iran

ABSTRACT: This paper presents the magnetic design of a 100MeV Separated Sector Medical Cyclotron (SSMC) as well as the optimization of the injection line magnets where will be used to transport the 14MeV proton beam from the Azimuthally Varying Field (AVF) cyclotron to the SSMC. The study demonstrates that the isochronous magnetic field with a tolerance around $10^{-4}$ T can be obtained all along prescribed path by placing three sets of correction trim coils in the pole tips. The result of Betatron oscillations is sufficient to verify that the focusing forces of the magnetic field can hold particles close to the median plane of the magnet. In order to inject the beam by minimum loss from a 14MeV Cyclotron to SSMC, it is essential for the beam to fit into the acceptance ellipse of the separate sector cyclotron. The characteristics of the beam injection system is calculated and optimized by Trace-3D beam dynamic code.

# 1. Introduction

In accelerator technology, magnets work like the highway for the charged particles to travel along the specific path and directions [1]. Optimum design and fabrication of all types of magnets in a typical accelerator can guarantee the final beam quality. One of the magnets applications is guiding the beam from one accelerator to the other which is inevitable in high energy cyclotrons or synchrotrons [2-5]. There are some medium-scale and baby-cyclotrons in Iran for radio-pharmacy production purposes which are used in cancer diagnostic applications. Moreover, at the moment, the construction phase of a 10 MeV proton compact cyclotron is on the progress [6,7]. This cyclotron is for production of short life positron-emitting isotopes which is used in positron emission tomography (PET) systems [8]. However higher energies are very useful for proton therapy. Thanks to the Bragg-peak at the end of proton range in a tissue, proton therapy has a very unique therapeutic properties which is the only choice to treatment some kinds of tumors like *Ocular* Melanoma, or malignant tumor in very sensitive organs such as the brain and spinal cord [9]. Unfortunately, no therapeutic proton accelerator exists in Iran and there is a wide demand for this kind of facility for cancer treatment in the region. As a more serious step in the design and development of cyclotrons to overcome increasing demand for high energy medical accelerators, we intend to design a separated sector medical cyclotron (SSMC) which will be capable of accelerating proton up to 100 MeV energy (See Fig.6 & 16 for layout of SSMC) [10,11]. This machine is a single particle fixed-energy accelerator and adopts four separated sector electromagnets which are made of steel-1010 and 8 normal conductor coils. In our design, 24 normal correction trim coils are used to obtain isochronism condition through the entire acceleration path. The empty space between magnet sectors provides the possibility of mounting of two RF cavities, vacuum pumps, injection line and the other subsystems. Design of separated sector cyclotron is complicated by the fact that particles cannot be accelerated from very low energy to high energy due to limitation in the magnet design [12]. To overcome this problem, the extracted beams from a lower energy accelerator such as cyclotron, linac or van de graff, are radially or axially injected into the median plane of the separated sector cyclotron. The injection line design for each case depends on their specification and have its own challenges.

This paper includes three main sections. Section 2 presents the design and optimization of the magnetic structure of a 100 MeV SSMC [13]. 2-D simulation by POISSON code [14] was started in order to determine basic magnet dimensions, coil shape, and current settings. After determining the essential parameters of the magnet such as the magnets dimensions, magnetic field spatial map, number of sectors in SSMC, harmonic number, RF frequency, hill and valley gaps (Hills are with higher magnetics' field as valleys are with lower magnetics' field, See Fig. 6) and angle of the pole, an initial simple model is established. Then, iterative process was accomplished in order to obtain an isochronous magnetic field by utilizing 24 normal correction trim coils using OPERA3D software [15]. The map of magnetic field is imported to CYCLON code [16] to calculate the integrated phase slip and Betatron oscillation [17]. The results of this section will be used to fabricate the magnets of the SSMC in the near future.

In section 3, we describe the optimization procedure of the injection line for radial injection of 14 MeV proton to 100 MeV SSMC by using 11 quadrupoles, two bending magnet and one RF cavity is presented and discussed. The characteristics of the beam injection system including the length of the system, number of quadrupoles and their gradients, distance between them, radius and angle of bending magnets, and voltage of the RF cavity is calculated by Trace-3D code [18].

## 2. Design of SSMC magnets

Prior to 2-D simulation, some basic parameters of the magnet have been estimated. The maximum kinetic energy of the accelerated particles which determines the magnetic rigidity at extraction radius can be expressed by equation (1) [19].

$$B.\rho = \frac{\sqrt{T^2 + 2TE_0}}{300Z} \tag{1}$$

Where, B is the average isochronous magnetic field (Tesla), ρ is the bending radius (m), Z is the charge state of the accelerated particle, T is the kinetic energy (MeV) and $E_0$ is the rest mass of the particle (MeV). According to the magnetic rigidity, the extraction radius and the magnetic field at extraction radius were figured out 1.49 m and 1.79 Tesla, respectively. Four-fold configuration for sector magnets with the angle of 50 degrees was selected to avoid entering to resonance zones [20-23].
The maximum value for the axial gap between poles was chosen 4 cm. This small gap not only reduces the number of Ampere-turns, but it also improves the vertical focusing of the magnet. By decreasing the hill gap of the magnet from 4 cm at injection radius to 3.4 cm at extraction radius, increasing magnetic field is obtained. The theoretical isochronous magnetic field [24,25] at the hill of the magnet versus radius of the pole in different RF frequencies is shown in Fig. 1. As illustrated in Fig. 1, the greater RF frequency leads to greater magnetic field on sectors at the same radius.

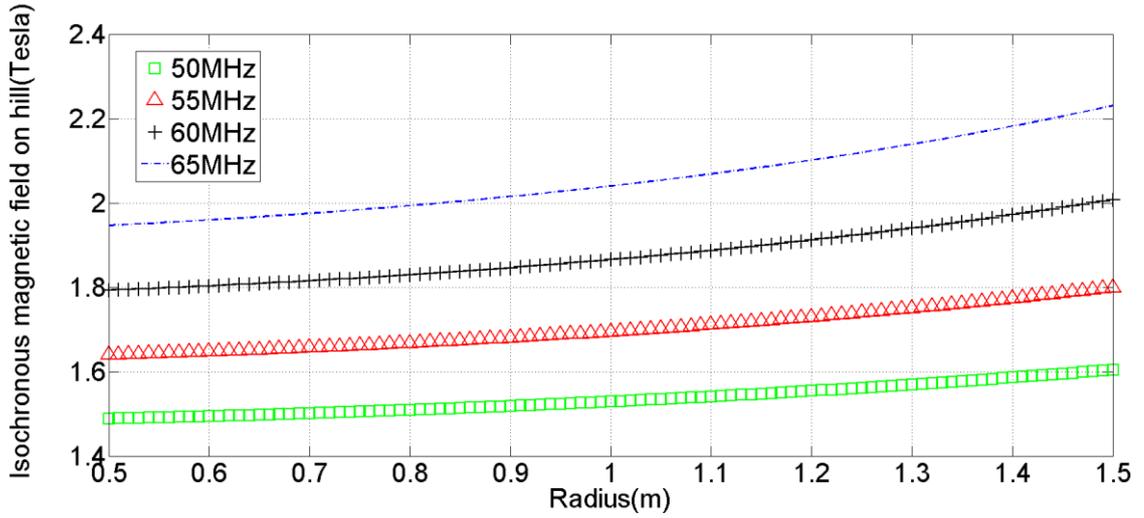

**Figure 1.** Isochronous magnetic field at hills versus radius of the pole.

Since the iron will be magnetically saturated at 1.9 Tesla, in order to be far from iron saturation, 55MHz was selected as an appropriate RF frequency for the designed magnet [26,27]. The other important parameters of the magnet were obtained by simulation and the final key parameters are listed in Table 1.

**Table 1.** Main parameters of the SSMC sector magnet.

| Parameter | Value |
| --- | --- |
| Injection / Extraction energy | 14 / 100 MeV |
| Ion type (pre accelerator) | $H^-$ |
| Ion type (Separated Sector Cyclotron) | $H^+$ |
| Number of sectors | 4 |
| Open angle of the sector | 50° |
| Pole radius | 165cm |
| Height of the magnet | 280cm |
| Inner/ outer radius | 40cm/300cm |
| Min/max of Hill gap | 34mm/40mm |
| Maximum magnetic field on hill | 1.9 Tesla |
| Magnet weight | 305t |
| RF frequency | 55MHz |
| Accelerating voltage | 250 kV |
| Harmonic number | 4 |
| Number of Dees | 2 |
| Number of Ampere-turns (main coils) | 26 kAT |
| Number of Ampere-turns (trim coils) | 0.2, 0.33, 0.25kAT |

### 2.1 Two Dimensional Modelling of SSMC Magnet

After considering all theoretical aspects, we have done 2-D simulation by POISSON code in order to determine basic magnet dimensions, coil shape, and current settings [14].

The execution of automesh in POISSON code provides the possibility of generating fine triangular mesh which can easily fits to boundaries of different geometry. This capability is useful to study details or do local modifications, which produce local effects. Iterative process was done in order to optimize magnet profile based on the specifications of the theoretical isochronous magnetic field [28-30]. Fig. 2 presents the 2-D field distribution on the vertical symmetry plane on the hill of the magnet.

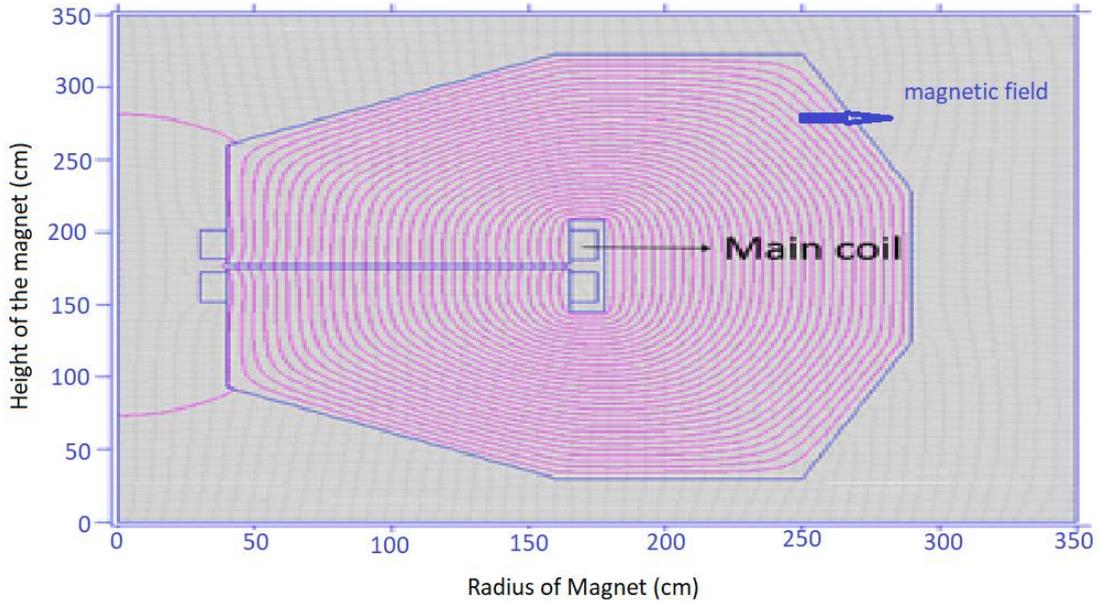

**Figure 2.** Equal-vector-potential field lines in vertical cross section of the magnet simulated by POISSON code.

As revealed in Fig. 3, the simulated magnetic field is not in consistent with the theoretical one (which are calculated by referring to the reference [24]) at initial and final radii of the magnet. In order to satisfy the isochronous condition, some trim coils should be implied to the magnetic structure.

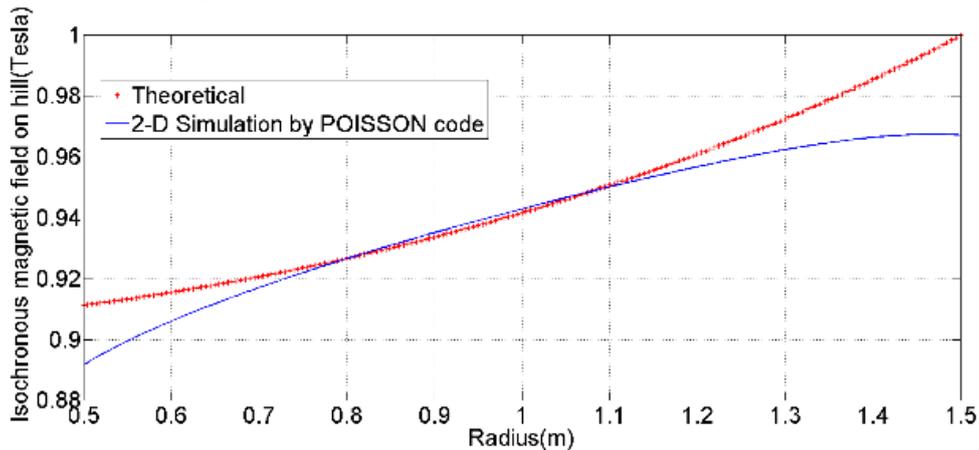

**Figure 3.** Deviation of simulated isochronous magnetic field on hill versus radius of the hill against the theoretical model.

By successive iterations we could estimate the trim coils current setting and their positions. It was found that the required magnetic field at the hill of the magnet sectors can be obtained by placing 3 sets of correction trim coils in the pole tips. The return path direction of magnetic field for inner trim coils is clockwise and for outer trim coils is counterclockwise [31,32]. Fig. 4 shows the position of correction trim coils and their magnetic field distribution in the absence of main coils. In order to ensure the stability of the magnetic field in vertical direction, trim coils were fixed inside the poles rather than locating them on the surface of the poles. Some parameters such as trim coil positions, dimensions and current settings were optimized in order to achieve the required isochronous magnetic field.

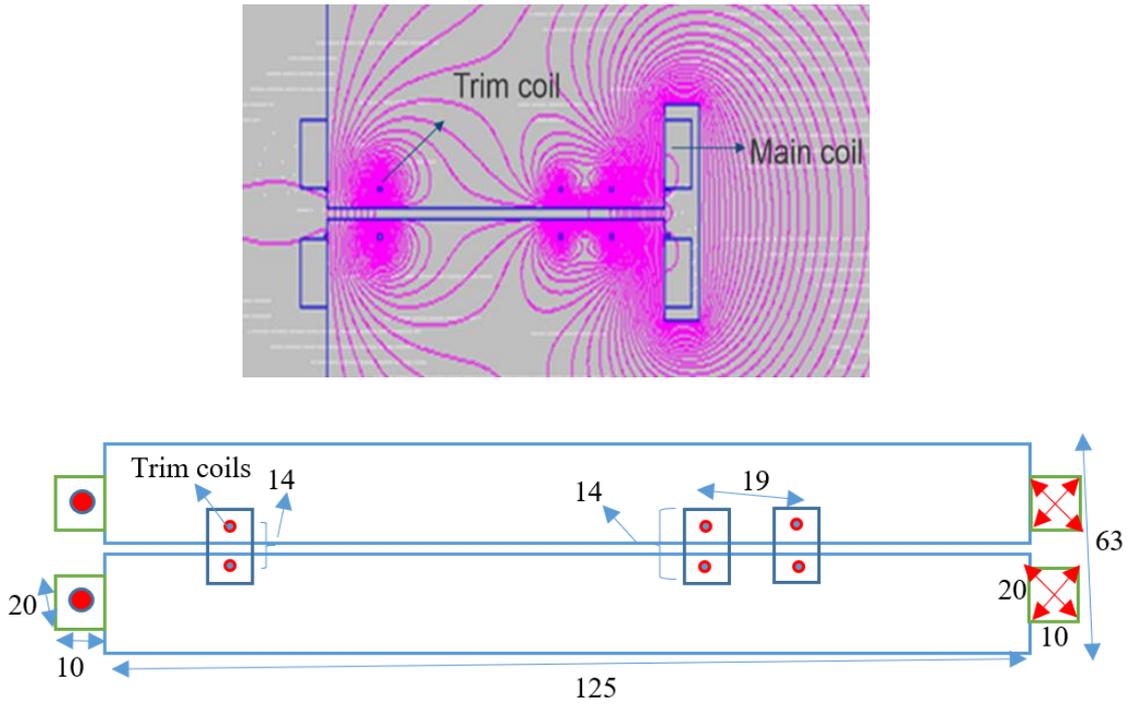

**Figure 4.** The vector potential lines due to trim coils in the absence of field due to main coil. (Below) shows sizes and distances of these coils from each other.

Fig. 5 shows the deviation of the final 2-D simulation relative to theoretical results. As illustrated in Fig. 5, an acceptable approximation to the isochronous magnetic field is obtained with uncertainties less than 15 Gauss for most of the radii. This model gives an approximation to 3-D model that will later be constructed by OPERA3D software [15].

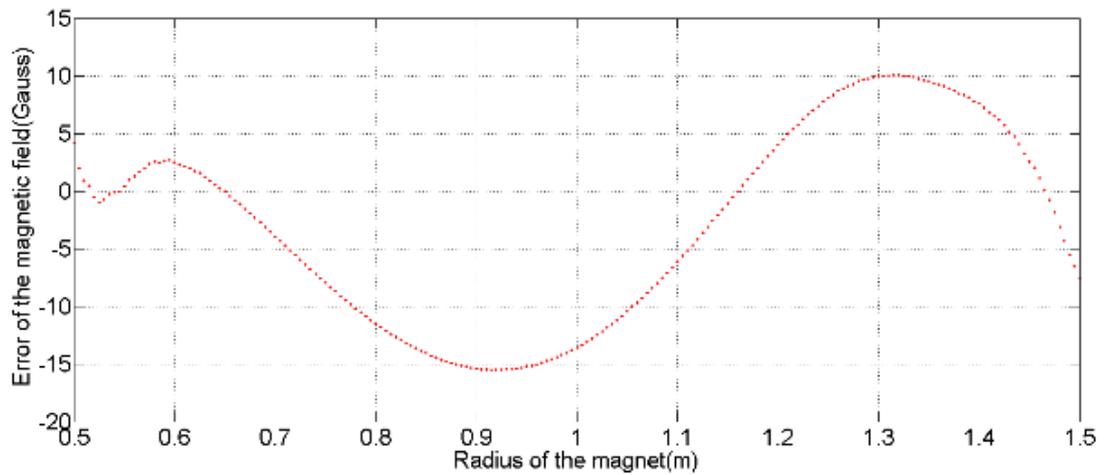
**Figure 5.** Field error versus radius of the pole.

### 2.2 Three Dimensional Modelling of SSMC Magnets

In this section we explain 3-D modelling of SSMC sectors. For doing an acuurate simulation, the B-H curve of the magnet's material provided by the factory and real geormetry of magnets were imported to the OPERA3D software [15].
Tetrahedral mesh is adopted to obtain a consistent model with irregular cells in geometry [33-36]. Smaller local mesh for median plane is used to increase the precision of simulation. Fig. 6 shows the whole model of the magnet which is meshed by 6,532,128 tetrahedral elements.

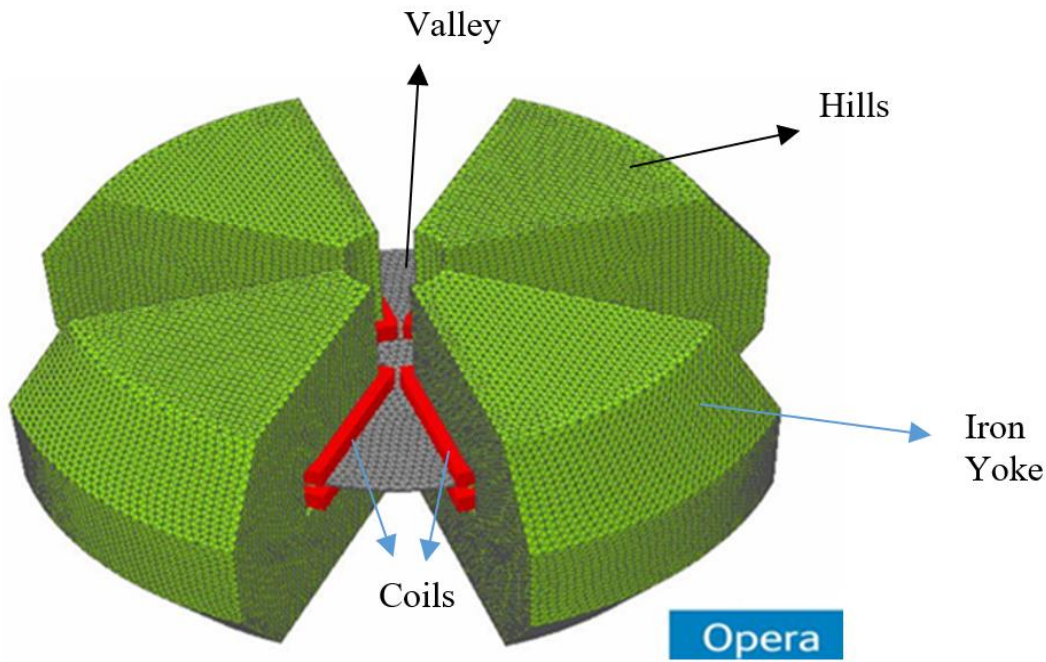

**Figure 6.** OPERA 3-D tetrahedral meshed model.

Hard edge shimming method is used to make the error of the designed magnetic field approaches zero as much as possible. After the modification, the results of calculation are studied by OPERA-3D post processor and errors in the isochronous magnetic field are well controlled to achieve a magnetic field with a tolerance around $10^{-4}$ T [15,37,38]. Fig. 7 shows the error of the 3-D designed magnetic field versus radius of the magnet. Fig. 8 shows the final profile of the pole.

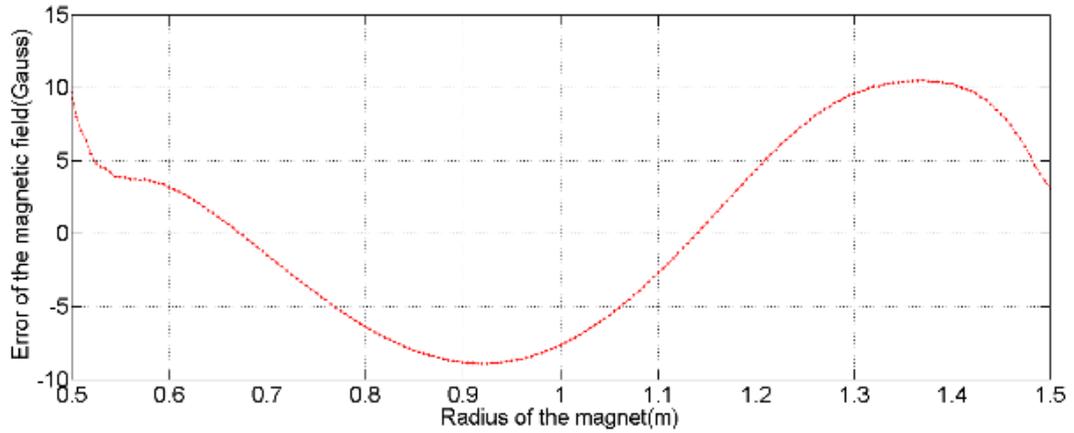

**Figure 7.** Error of the designed magnetic field versus radius of the magnet simulated by OPERA-3D.

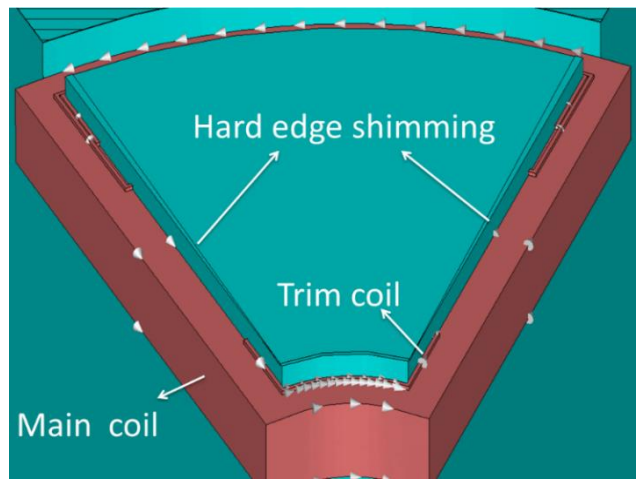

**Figure 8.** Final profile of the pole

We used calculated magnetic fields by OPERA-3D to calculate the trajectories and equilibrium orbits and the phase slip of the particles [14,38]. Fig. 9 shows the OPERA 3-D post processing results of the magnetic field on mid-gap of the magnet.

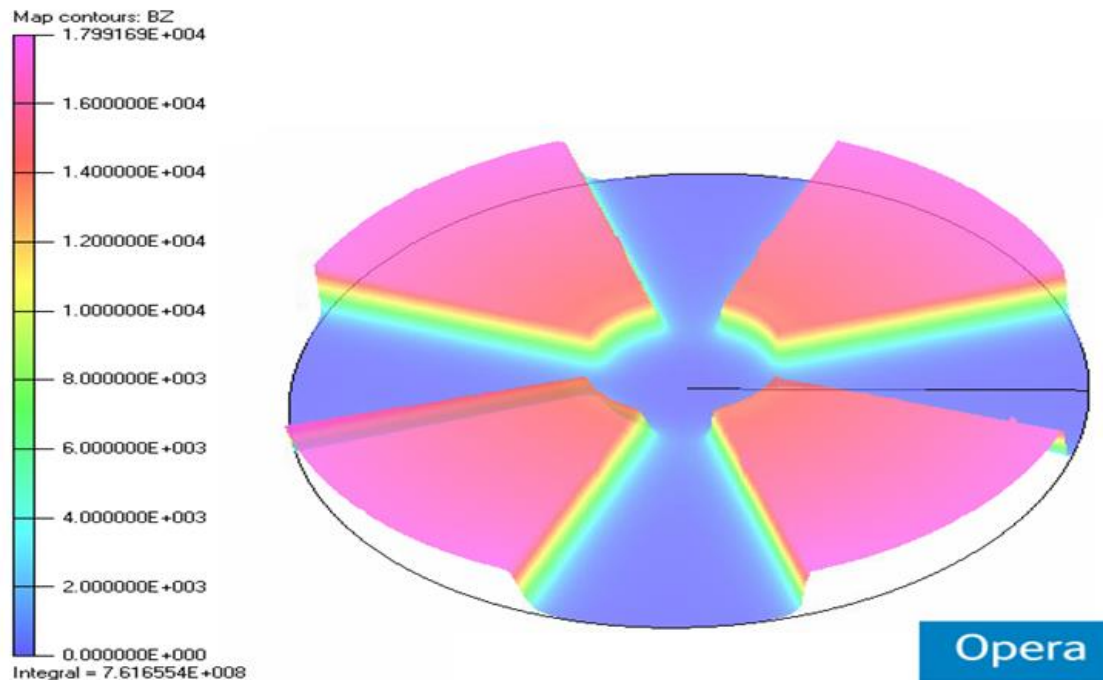

**Figure 9.** Spatial map of magnetic field on the median plane of the magnet.

When the isochronous magnetic field with the 4$^{th}$ harmonic RF cavity have been created, closed orbits of the particles up to energy at extraction radius were found. As illustrated in Fig. 10 the results show the appropriate trajectories for particles in mid-gap of the proposed magnet.

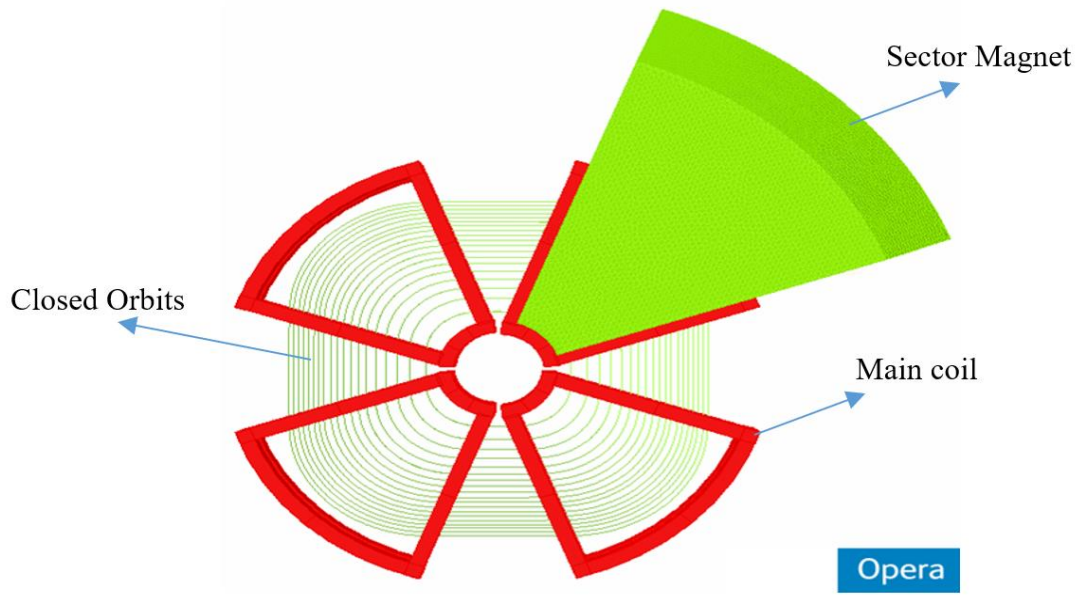

**Figure 10.** Closed orbits of the magnet.

Fig. 11 shows the integrated phase slip versus radius of the magnet which is calculated by CYCLONE code [16]. The acceptable values of the integrated phase slip of the particles besides the proper beam dynamics of the closed orbits are sufficient to confirm that the designed magnetic field is isochronous [39-41].

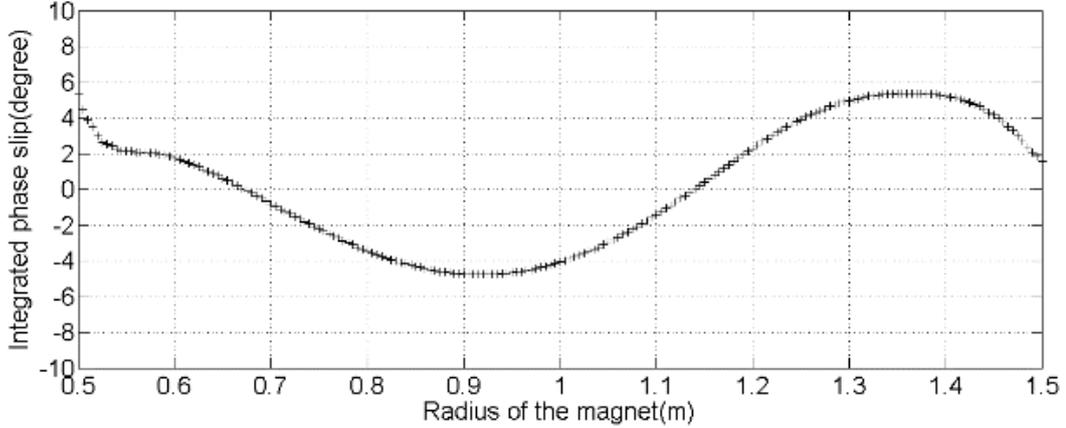

**Figure 11.** Integrated phase slip versus radius of the magnet.

In the absence of focusing forces, Betatron oscillations can capture particles with large oscillation amplitude and bring them in collision with the vacuum chamber. The characteristic properties of the magnetic field distribution in the acceleration gap are responsible for all kinds of resonances such as integer, half integer and linear coupling resonances. For working points to be far away from all types of dangerous resonances, the magnetic field in the acceleration gap should ensure strong axial and radial focusing of the beam [41].

The radial and axial Betatron tune, $\upsilon_r$ and $\upsilon_z$, depends on multiple parameters of the magnet which can be approximately expressed by [41],

$$\upsilon_z^2 = 1 - \gamma^2 + \left[ N^2 / (N^2 - 1) \right] F(1 + \tan^2 \zeta) \quad (2)$$

$$\upsilon_r^2 = \gamma^2 + \left[ 3N^2 / (N^2 - 1) \right](N^2 - 4) F(1 + \tan^2 \zeta) \quad (3)$$

Where $\zeta$ is the spiral angle and is zero for the proposed non-spiral magnet and $N$, $F$, $\gamma$ are number of sectors, field flutter and relativistic factor, respectively. The values of Betatron oscillation frequencies have to be positive. It is necessary to change the model and to do all initial calculations again if the value of $\upsilon_z$ is negative or $\upsilon_r$ is smaller than 1 [42-44].

Calculation of Betatron oscillation has been done by importing OPERA3D magnetic field map to CYCLONE code. Fig. 12 shows the values of Betatron oscillation versus radius of the magnet pole. As illustrated, both $\upsilon_z$ and $\upsilon_r$ are positive and $\upsilon_r$ is higher than 1. So, the magnetic field distribution can provide horizontal and vertical focusing forces which can hold particles close to the median plane of the magnet.

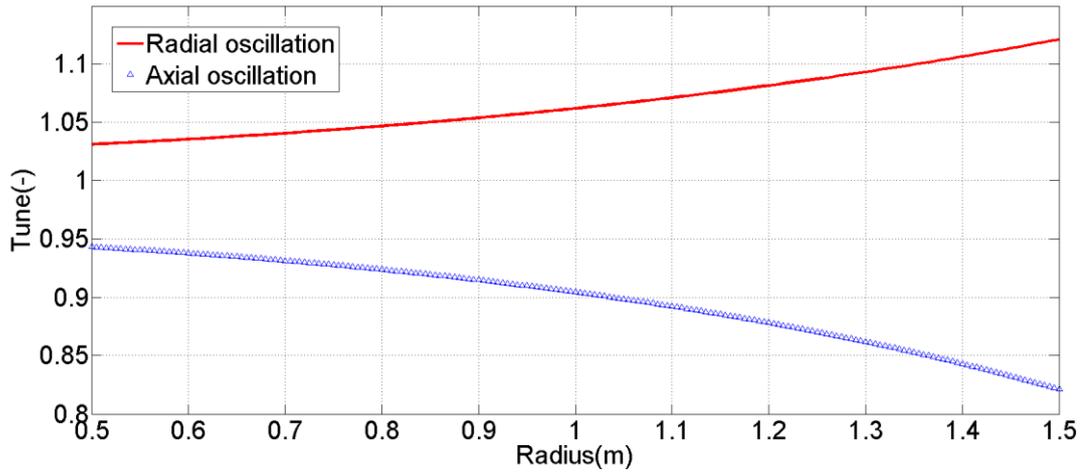
**Figure 12.** Axial and radial tune versus radius of the magnet pole.

Fig. 13 shows the resonance diagram up to third order approximation with $|k|+|l| \leq 3$ for designed super-periodic magnet [44]. The operating points of the designed magnet in the radial and axial planes are displayed by a curve on the tune diagram. The results verify that the operating points do not have any intersection with dangerous resonance lines. So, the magnetic field distribution does not need revision from this aspect.

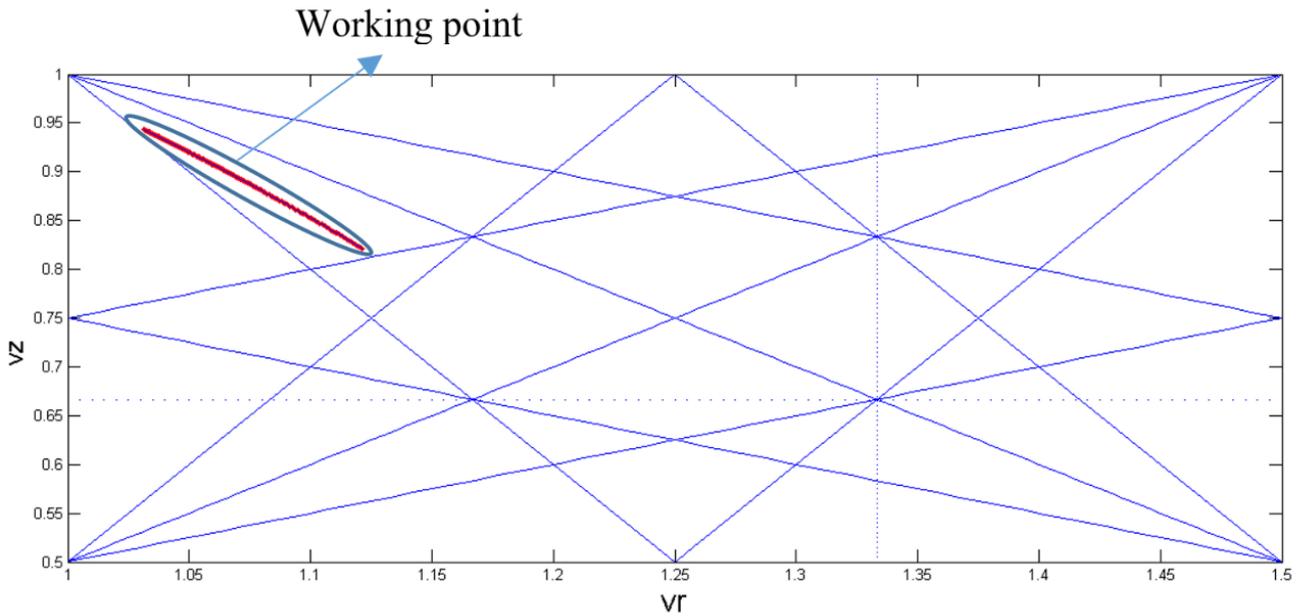
**Figure 13.** Resonance diagram up to 3$^{rd}$ order and working point.

## 3. Optimization and specification of injection line magnets

The main goal of optimization of the injection line is to find the line components parameters including length of the system, number of quadrupoles and their gradients, distance between them, radius, angle and peak field of bending magnets, as well as
–10–

voltage of the RF cavity. Trace3D code try to change the initial magnet parameters until the beam dynamic parameters at the end of the beam line is adopted to the desired values [18].

In our case, the total length of the injection line from the exit of the 14 MeV cyclotron to the centre of SSMC is 17 meter. The Twiss parameters and other beam parameters of 14 MeV cyclotron at the entrance of injection line is shown in Table 2. These parameters should be measured experimentally

**Table 2.** Main parameters of the proton beam at the entrance of the injection line.

| Parameter | Unite | Horizontal Value | Vertical Value |
|---|---|---|---|
| Emittance | $\pi.mm.mrad$ | 250.1 | 40.3 |
| $\alpha$ | - | -6.168 | 1.123 |
| $\beta$ | mm/mrad | 2.77 | 0.954 |
| $\gamma$ | mrad/mm | 14.1 | 2.38 |
| Beam energy | MeV | 14 | |
| Beam current | mA | 2 | |

As it was mentioned, in order to inject the beam by minimum loss, it is essential for the beam to fit into the acceptance ellipse of the SSMC. The beam dynamic parameters at the end of injection line should be fulfill these criteria. The size of the proton beam at the end of the injection line is determined by its Beta function value. For high injection efficiency, usually, the acceptance ellipse is considered 7 times of injected beam emittance. It is possible to calculate the acceptance ellipse of the SSMC using equation (4) and (5), assuming that the lattice function (alpha) is zero for the acceptance ellipse [42].

$$7\varepsilon_x = \frac{(d - D(s))^2}{(\beta_x)_{\min}} \Rightarrow (\beta_x)_{\min} = 0.07 \, mm/mrad \quad (4)$$

$$7\varepsilon_y = \frac{d^2}{(\beta_y)_{\min}} \Rightarrow (\beta_y)_{\min} = 0.45 \, mm/mrad \quad (5)$$

In these relations, $d$ is gap value of the dipole of the sector (mm), D(s) is the dispersion function which is zero in the ideal lattice. $\varepsilon_x$, $\varepsilon_y$ are the horizontal and vertical normalized emittance, respectively. These values are related to transversal direction. The longitudinal direction (z-direction) parameters are the energy and phase of the particles. There is an analytical relation between the energy deviation acceptance and the phase of the particle relative to the RF voltage in cyclotron as follow [42],

$$\Delta E = \sqrt{\frac{2G\beta^2 eU_0 E\left[\cos(\psi_0) + (\psi_0 - \pi/2)\sin(\psi_0)\right]}{\pi h\left(\alpha - \frac{1}{\gamma^2}\right)}} \quad (6)$$

Where, G=-1 for $0 < \psi_0 < \pi/2$ and G=+1 for $\pi/2 < \psi_0 < \pi$, $\beta, \gamma$ are relativistic factors, $U_0$ is the RF voltage amplitude (V), $E$ is the particle energy (MeV), $h$ is the harmonic of the cavity, $e$ is the electron charge and $\alpha$ is the momentum compaction factor which it is equal to 0.029 for our 14 MeV cyclotron [45-47]. It is useful to plot the energy acceptance of the SSMC versus the phase deviation which is shown in Fig. 14. This plot is very useful to find the energy acceptance of the SSMC for specific phase deviation and is important to design the end section of the injection line.

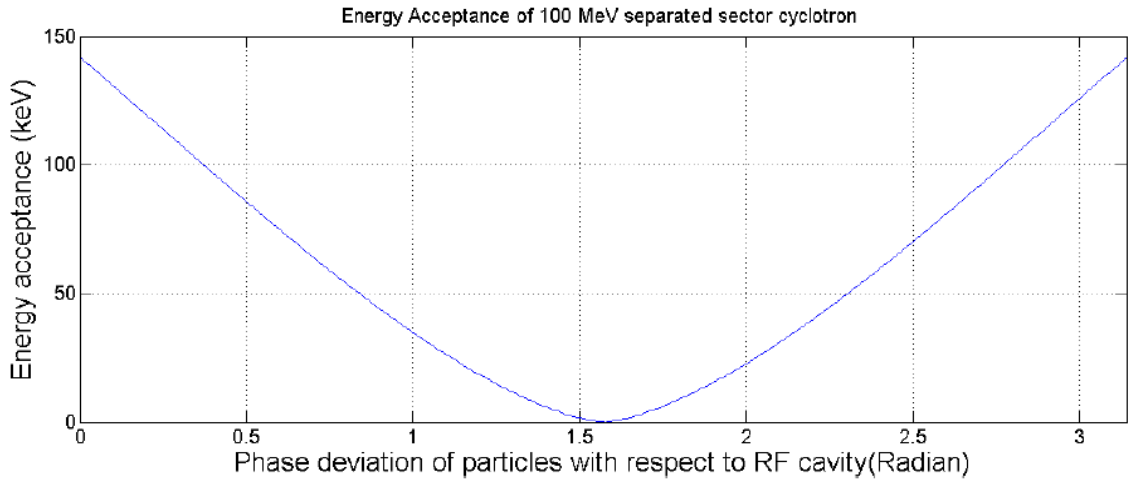

**Figure 14.** Energy acceptance of the SSMC versus the phase deviation.

The optimization procedure is started by calculation of proton beam parameters at extraction point of 14 MeV cyclotron in free space tube. Fig. 15 shows the beta function along the free injection tube. As we can see, after almost 50 cm the beam divergence will be high, and the beam will be near to the tube wall and it is necessary to control its trajectory. The well-known FODO lattice is used to control the beam divergence [46-47]. FODO is a pair of quadrupole magnets which focus the beam in two directions. This approach is implied along the tube line to ensure that the proton beam is reached to end of the injection line.

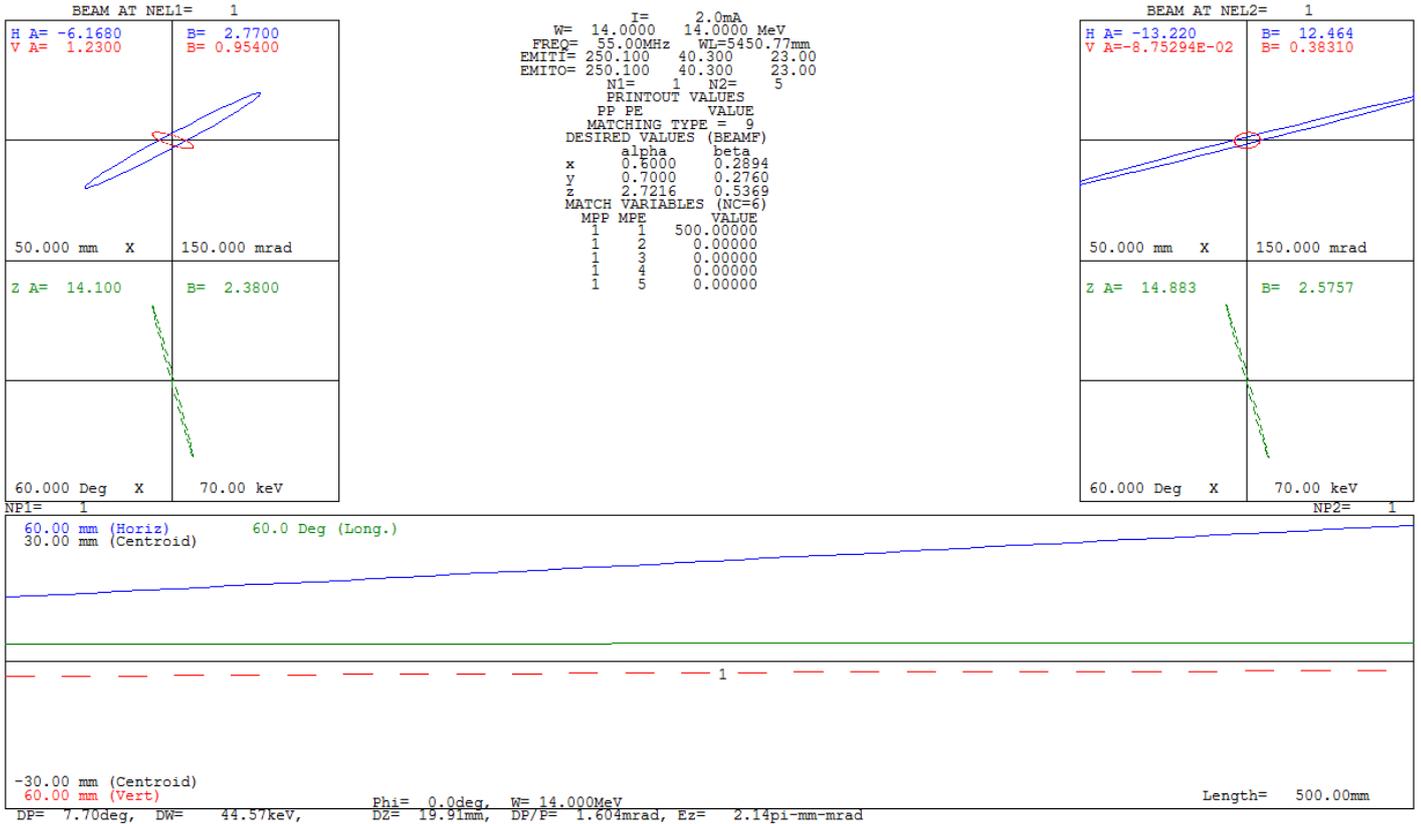

**Figure 15.** Trace-3d calculation of Beta function along the free injection tube before adding quadrupoles.

After some simulation steps, the final arrangement of the magnets in the injection line will be determined at this step of optimization. It is obvious that the Beta functions of the beam satisfy the SSMC acceptance ellipse but the energy and phase deviation at the end of the injection line are 54.47 keV and 19.4 degree, respectively. According to the Fig.14 for 19.4 degree of phase deviation the energy acceptance is 100 keV which is above the energy deviation of injected beam. However, the value of 54.47 keV for energy deviation is very high to inject the beam into the SSMC. The CYCLONE calculation shows that the energy deviation of the beam after each complete rotation at SSMC will be worsen [16, 24]. We find that in order to reach an appropriate value for energy resolution at the exit of SSMC, the energy deviation of the injected beam must be below 4 keV. For this reason, it is a good idea to use an RF cavity at the end of the injection line. For some technical reason and due to shortage of enough space, the RF cavity is mounted before the final triplet. According to the Trace calculations, the final energy deviation after using the RF cavity is 1.4 keV.

The final optimized parameters of the quadrupoles, bending magnets and the RF cavity are derived by TRACE3D and listed in Table 3 [18]. The final layout of the SSMC and injection line is shown in Fig 16.

**Table 3.** Optimized parameters of the injection lines magnets.

| Parameter | Value |
|---|---|
| Number of Quadrupoles | 11 |
| Quadrupoles Gradients | 14 T/m |
| Quadrupoles Thickness | 15 cm |
| Bending magnet field | 1.44 T (magnet 1) 1.64 T (magnet 2) |
| RF cavity frequency/Phase/Voltage | 55 MHz /- $\pi$ /2 /178 kV |

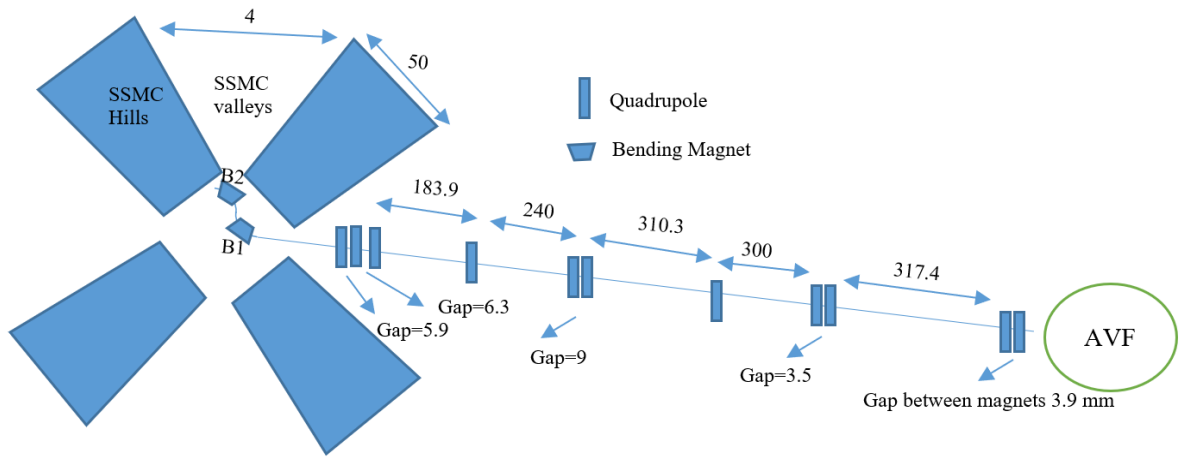

**Figure 16.** Final layout of the SSMC and injection line (units are in mm).

## 4. Conclusion

Proton therapy facilities are growing rapidly around the world. This paper presents the magnetic design of a separated sector cyclotron and its injection line as the first proton therapy facility in Iran. The simulations have been performed to get the isochronous magnetic field for a 100 MeV separated sector cyclotron. The preliminary design of the sector magnets is obtained by POISSON code. Modelling of correction trim coils demonstrates that the isochronous magnetic field can be established within high rang of precision all along acceleration path in the median plane of the magnet. When the key parameters of the magnet have been specified, the simulation was followed in complete details by OPERA3D software to find optimal shape for sectors. The agreement between theoretical and simulated isochronous magnetic field by OPERA3D software is found to be excellent. The closed orbits of the particles show the appropriate trajectories for particles in the median plane of the magnet. Moreover, calculation of Betatron oscillations by CYCLONE code not only confirms the focusing properties of the magnet but it also indicates that working points are far away from dangerous resonance zones. The injection line was optimized by Trace 3D code in order to transport the beam from a 14 MeV AVF cyclotron to separated sector cyclotron. An RF cavity was inserted at the end of the injection line to improve the energy deviation of the beam to 1.4 keV. The results of the simulation have been used for fabrication of the first large cyclotron facility in the region with medical purposes.